\documentclass[showpacs,preprintnumbers,amsmath,amssymb,floatfix]{revtex4-1}

\usepackage{color}
\usepackage{graphicx}
\usepackage{subfig}
\usepackage{dcolumn}
\usepackage{bm}
\usepackage{enumerate}
\usepackage{setspace}
\usepackage{listings}

\normalsize

\begin{document}

\title{Radiative and atomic properties of C and CH plasmas in the warm-dense matter regime}

\author{T.-G.~Lee,$^{1}$ M.~Busquet,$^2$ M.~Klapisch,$^3$ J.~W.~Bates,$^1$ A.~J.~Schmitt,$^1$ S.~X.~Hu,$^4$ and J.~Giuliani$^1$}

\affiliation{$^1$Plasma Physics Division, U.S. Naval Research Laboratory, Washington, D.C. 20375}
\affiliation{$^2$Research Support Instruments, 4325 Forbes Blvd, Lanham, MD 20706}
\affiliation{$^3$Syntek Technologies Inc., 801 N Quincy St Ste 610, Arlington, VA 22203}
\affiliation{$^4$Laboratory for Laser Energetics, University of Rochester, Rochester, New York, 14623}

\def\bb#1{\hbox{\boldmath${#1}$}}
\def\pbar {\bar p}
\setstretch{1.5}   

\begin{abstract}
A theoretical model based on the method of super transition arrays (STA) is used to compute the emissivities, 
opacities and average ionization states of carbon (C) and polystyrene (CH) plasmas in the warm-dense matter 
regime in which the coupling constant varies between 0.02 to 2.0. The accuracy of results of STA calculations 
is assessed by benchmarking against the available experimental data and results obtained 
using other theoretical methods, assuming that a state of local thermodynamic equilibrium exists in the plasma.  
In the case of a carbon plasma, the STA method 
yields spectral features that are in reasonably-good agreement with Dirac-Fock and Hartree-Fock-Slater theories; 
in the case of CH, we find that STA-derived opacities are very similar to those derived using 
quantum-molecular-dynamics density-functional theory and Hartree-Fock method down to 
plasma temperature of about 20 eV. Our calculations also compare favorably with available 
experimental measurements of Gamboa {\it et al} [High Energy Density Phys. {\bf 11}, 75 (2014)] 
of the plasma temperature and average ionization state behind a blast wave in a pure carbon foam. 
Although the STA-computed average-ionization charge state in the rarefaction region 
appears to be lower than the experimental data, it is within the experimental uncertainty and 
the discrepancy is nevertheless consistent with results reported using an atomic kinetic model. 
In addition, we further predict the temperature dependence of average ionization states of 
CH plasma in the same temperature range as for the carbon plasma. 

\end{abstract}

\maketitle


\section{Introduction}
The development of intense and energetic lasers over the last several decades has 
given rise to a variety of important advances in the field of intense x-ray pulses \cite{A5}, inertial confinement fusion \cite{A6,A7} and
laser-driven nuclear phenomena \cite{A8}. Such lasers are also capable of creating macroscopic 
samples of warm-dense matter (WDM) \cite{wdm1}, which is of great interest and relevance to 
astrophysical research and potential industrial applications. 
The WDM regime lies between condensed matter state and hot plasma, and is characterized 
by a temperature range within an order of 1 eV 
and by an ionic density $\rho_i$ of order unity in the coupling parameter
$\Gamma = (4\pi \rho_i/3)^{1/3}({\rm \bar Z}e)^2/k_BT$
and the electron degeneracy parameter $\eta$ = $k_BT/E_F$ ($E_F$ is the Fermi energies 
and $k_B$ is the Boltzmann constant). For ions other than hydrogen, however, actually characterizing
the properties of plasma in the WDM regime \cite{wdm2} is a formidable task.    

Experimentally, intense and spectrally tunable x-ray light sources 
such as the Linac Coherent Light Source facility at SLAC and ORION provide 
a valuable diagnostics tool for studying WDM \cite{xray1,xray2,xray3,xray4,xray5,xray6}. 
These intense light sources can be used to probe a WDM sample and provide information 
on its underlying structure, charge states and opacity. WDM created by short laser 
pulses is however, highly transient with a time-scale on the order of a nanosecond 
and likely requires diagnostic resolution times that are on the order of tens of picoseconds or shorter. 
This immediately sets restrictive limitations for most present day experiments. 
As a result, accurate theoretical and computational models of the atomic and radiative properties of WDM 
must be developed concurrently with experimental efforts and play a central role in helping 
to advance our understanding of the physics of this WDM regime.   

There are several theoretical and computational approaches for evaluating the radiative 
and atomic properties of plasmas in a state of either local thermodynamic equilibrium (LTE) 
or non-local thermodynamic equilibrium (NLTE). For dense plasmas composed of 
transition elements with high nuclear charge, methods based on the detailed line accounting 
(DLA) framework \cite{STA1} (which computes a fully resolved spectral lines 
of all configuration-to-configuration transition arrays) are 
computationally prohibitive due to the complex interaction among configurations 
containing several thousand of states with billions of transitions.
Atomic physics models based on the unresolved transition arrays (UTA) 
framework \cite{UTA} (which assumes that all lines in the spectrum of each 
configuration-to-configuration transition arrays merge into a single effective 
line of a Gaussian shape) are somewhat more practical and are often employed \cite{STA1}. 
But in many of those cases, the number that of relevant UTAs is still too large to 
be computed in a short period of time. A more efficient approach based on the  
methods of super transition arrays (STA) was introduced and pioneered by 
Bar-Shalom {\it et al.} \cite{Avi89} provides a practical and less time-consuming alternative.

The STA method \cite{Avi89,Avi94,Avi95,Avi96,Avi97,Avi99,PRTA1,PRTA2,Yair16,STA1} 
models electronic transitions via a statistical partition function (PF) formulation by gathering 
ordinary electronic shells into supershells, electronic configurations into superconfigurations (SCs) 
and groups transition arrays into STAs. The supershells and superconfigurations
can be dynamically defined and refined iteratively. The averaged atomic quantities such as energies 
and widths of the transition arrays, and occupation numbers are then determined through the computation 
of the PF of the supershells, which are populated in all possible ways according to the Pauli exclusion principle. 
However, the model assumes the followings \cite{Yair16}: (i) The plasma is in local thermodynamic equilibrium. 
(ii) All configurations forming a superconfiguration 
are subject to a common mean central-potential (optimized for the superconfiguration) 
with the same set of one-particle solutions. (iii) A high-temperature approximation is valid,
which in this study implies that the plasma temperature must be approximately greater 
than the spread of the energies of configurations within a superconfiguration. In this limit,
the population of each configuration in the superconfiguration is proportional to the partition function. (iv) 
The spectra of all UTAs which form an STA merge into a
single Gaussian function. 

These assumptions immediately pose the following questions: 
If STA is known to be a high-temperature model, 
how will it perform in the low-temperature regime? 
Will it lead to a significant loss of accuracy? 
Although STA was formulated specifically for simulating mid- to high-Z plasmas, 
is it still valid for low-Z plasmas? Will it lead to a significant loss of spectral features? 
Here, we wish to address these questions by examining the radiative and atomic properties of C and CH 
plasmas in the warm-dense regime. Plasma species containing carbon and CH are not only 
a litmus test of the validity of the STA model for the low-Z element, more importantly, 
the CH polystyrene is a material of choice that features prominently in many 
inertial-confinement-fusion (ICF) target designs. In ICF, high-intensity laser light or x-rays 
are used to implode a capsule of cryogenic deuterium and tritium by illuminating a 
spherical ``ablator" shell surrounding the fuel. CH is a popular choice for the ablator 
material in an ICF capsule because it is inexpensive, easy and safe to manufacture 
and has good laser-absorption properties, high hydrodynamic efficiency and 
low radiation preheat.  In particular, to achieve high thermonuclear yield, 
a successful ICF target design must not only maximize the ablation pressure 
and mass ablation rate but also minimize the radiation emitted by the ablator, 
which can preheat the DT fuel, reducing compressibility and spoiling high yield. 
Because the former two processes scale as the ratio of the atomic mass number to nuclear charge, A/Z, 
whereas the latter scales simply as Z, CH provides a reasonable balance of performance as an ICF ablator \cite{expCH}. 
Because of this application, theoretical and computational efforts have been carried out 
at the University of Rochester - Laboratory for Laser Energetics to simulate the equation of 
state (EOS) and optical properties of polystyrene CH \cite{Hu0,Hu1,Hu2,Hu3} 
in a wide range of densities (0.1 to 100 g/cm$^3$) and temperatures (10$^3$ to 4$\times$10$^6$ K) 
using quantum molecular dynamics (QMD) simulations based on a finite temperature Kohn-Sham density 
functional theory. Numerical simulations based on the QMD method are widely used to model WDM phenomena 
and have been successfully applied to interpret experiments of expanded metals \cite{Clerouin} 
and ICF implosions \cite{ICF1,ICF2, ICF3}. The STA method, on the other hand, based on solving 
the relativistic Dirac equation, has its advantage as it enables statistical completeness of atomic 
structure for arbitrarily complex ions at a fractional computational cost of QMD simulations.       

In this contribution, we apply a super transition arrays (STA)
self-consistent model in LTE to study the emissivities, 
opacities and average ionization charge state $\rm\bar Z$ for carbon as well as for CH plasmas under 
warm-dense conditions with values of the coupling constant $\Gamma$ varying from 0.02 to 2.0. 
By evaluating the accuracy of the STA calculations in comparisons with other theoretical calculations 
and experimental measurements, we attempt to provide some answers to the above questions. 
In addition, the present study will allow us to gauge the capabilities and limitations of the present STA 
method so that improvements might be made in the future. 

The paper is organized as follows. In Sec. II, we describe the computational method.
In Sec. III, we present the STA results in comparisons with other available theoretical calculations 
and experimental data for carbon and CH plasmas. Finally, in Sec. IV, we summarize our results. 
Unless otherwise stated, all quantities are expressed in atomic units. 

\section{Computational method}

\subsection{STA formalism}

In high-density plasmas composed of medium- to high-Z elements, 
the number of populated electronic configurations can proliferate 
rapidly owing to the collisional excitations among various states. As a result, 
the corresponding number of electronic transitions among the configurations 
in each ionic stage can also grow prohibitively large, making the numerical computations intractable. 
A practical solution to this problem is to introduce a quantum statistical model of  ``super transition arrays". 
The model based upon superconfiguration accounting groups closely spaced electronic configurations together 
to form superconfiguration. It also judiciously defines these groups and serves as a bridge between 
the rather simple average atom (AA) model and the computationally-intensive detailed configuration accounting (DCA).  
Since the theoretical details of the method have been given elsewhere \cite{Avi89}, 
we outline the essential concepts behind the STA model in the remainder of the section.    
 
Starting with superconfigurations, we shall follow the notations used in \cite{Avi89}. 
A superconfiguration (SC) is constructed through 
collecting together neighboring (in energy) ordinary subshells
into ``supershells". For instance, by assigning supershell occupation number $Q_\sigma$ to each 
of the two supershells ($1s2s2p_{1/2}2p_{3/2}$)($3s3p_{1/2}3p_{3/2}$), we define 
a superconfiguration
\begin{eqnarray}
\Xi = \prod_\sigma(1s2s2p_{1/2}2p_{3/2})^{Q_\sigma}(3s3p_{1/2}3p_{3/2})^{Q_{\sigma+1}}\ldots,
\end{eqnarray}
where $\sigma$ denotes the supershells and $Q = \sum_\sigma Q_{\sigma}$ is the 
total electron occupation number. This implies that all the possible partitions of 
the $Q_{\sigma}$ electrons among the ordinary subshells $s \equiv \{n, l, j\}$    
are considered. An ordinary configuration $C$, on the other hand, 
is a special case of SC in which each supershell contains only one shell. 
Moreover, a reasonable number of SCs (typically a few hundred for medium-Z elements)
can contain a tremendous number of ordinary configurations. Although these SCs are 
loosely defined at the very beginning of the calculation, the precision for the spectrum can be improved by 
iteratively refinement of the SCs (i.e., subdivision of the supershells) until the resultant
spectrum converges to the DCA spectrum, and it is possible to calculate macroscopic thermodynamic variables 
such as pressure, internal energy, Helmholtz free energy, by averaging over a reduced 
number of SCs made up of large number supershells. To be clear, note that the 
thermodynamics is not fully defined by the electronic configurations, 
only the electronic part of the macroscopic thermodynamics quantities can be defined this way.

To evaluate the STA moments (i.e., the total intensity, the average energy, and variance) 
and SC average rates, one needs expressions for
the populations of the configurations and superconfigurations.
Assuming local thermodynamic equilibrium, all the configurations 
described by a SC $\Xi$, the population of any array of levels $i$ can 
be expressed through the Saha-Boltzmann's law, $U/U_i = N_i/N$, 
where $N$ is the total ionic number density, and $U_i$ and $U$ are the 
corresponding partition function. For example, for an ion with $Q$ number of electrons, the partition function
of the SC $\Xi$ can be expressed in terms of a summation over all
levels $i$ of all configurations $C$ is  
\begin{eqnarray}
U_\Xi = \sum_{C \in \Xi}  \sum_{i \in C} g_i e^{-(E^{(0)}_i+\delta E^{(1)}_{\Xi}-Q\mu)/kT}, 
\end{eqnarray}
where the sum of configuration statistical weight is given by the sum of
product of binomials
\begin{eqnarray}
\sum_{i \in C} g_i =  \sum_p \prod_{s \in C} \binom{g_s}{q_s}, 
\end{eqnarray}
and $g_s = 2 j + 1$, $q_s$ is the number of electrons in a subshell. We also have the relations
 $\sum_{s\in\sigma} q_s = Q_{\sigma}$ and $E^{(0)}_i = \sum_{s\in\sigma} q_s \epsilon_s$, where the latter quantity is the zeroth order energy. 
 The SC energies are given by the expression
 \begin{eqnarray}
E_{\Xi} = \delta E^{(1)}_{\Xi} + \sum_{\sigma} \sum_p \sum_{s \in \sigma}  q_s \epsilon_s, 
\end{eqnarray} 
where $\delta E^{(1)}_{\Xi}$ is the SC first order average energy 
 correction (see Eq.(86) in Ref.\cite{Avi97}), and $\epsilon_ s$ are the monoelectronic energy 
 and $ \sum_p$ means a summation over all the partition functions of the supershell occupation number, $Q_{\sigma}$.
It should be noted that it has been demonstrated in Ref.\cite{Avi89,Avi97},
that with a modified set of statistical weights and supershell occupation numbers, 
the STA moments and the non-LTE average transition rates can be expressed 
in terms of generalized partition functions. 

The STA computer code is based on an ion sphere (IS) model in a chemical picture \cite{inferno}. 
In this picture, the plasma is considered to consist of partially stripped ions of each element and 
free electrons shared among all ions. For a given atom with a set of the temperature and the density 
of interest, the algorithm first solves a finite-temperature Thomas-Fermi-Dirac equation \cite{TFD1,TFD2} 
and using the solutions in terms of relativistic wave functions to obtain the average ionization charge state $\rm\bar Z$ 
self-consistently with the free electrons in the ion sphere. A parametric potential has been used to describe the bound electrons as it
simplifies and yet captures the changes of the electronic potential for each ion stage \cite{KM71,KM77}. 
The STA algorithm, in the first iteration, starts by loosely defining very broad supershells, similar 
to that of the average atom model approach \cite{STA1}. The SCs are then constructed with these supershells. The moments of the STA 
transitions are then computed and the resulting spectra are obtained by adding up all the STA contributions. Then, in the next iteration, 
the supershells are split to optimize the corresponding SCs and 
this procedure repeats itself until the converged spectra are achieved. 
The potential for each SC is also progressively refined, the STA recomputed, 
and finally, the UTA moments are incorporated in the spectral-opacity calculation, 
as part of the obtaining accurate STAs' widths and energies. The bound-free and free-free 
transitions are also computed using the same potential. Ultimately, the convergence criteria of the STA-computed spectra 
using this iterative procedure is the DCA spectra.
 
\subsection{Mixture model}
There are several ways to compute an opacity for a mixture of chemical elements. 
Here, we use the word ``mixture" cautiously since our model is ``mechanical" and 
not a product of chemical reaction. Strictly speaking, we do not consider any quasi-molecular ion-ion 
intermingles or overlaps and/or condensed solid-state effects and all the elements are in the atomic or ionic state,  
as individually computed with the STA code. The mix model used in the present work has been described previously by 
Klapisch and Busquet \cite{mix}. 

The MIX algorithm requires an input file that describes the desired relative components 
of the elements and takes input data files generated by the STA computer program. 
It also reads an additional set of data files containing the desired photon groups (in bins) in order to create multi-group Rosseland and Planck opacities. 
The MIX algorithm extracts for each element a set of density values for each temperature point. It then numerically evaluates the partial densities for the 
chosen mix model in order to retrieve the corresponding data from the opacity database within the same temperature point. If necessary, 
it is capable of interpolating between two density points on the spectrally resolved opacities and then computes the group opacities. Once done, 
it moves on to the next temperature and repeats the process.  

\begin{figure}[!htp]
\centering
\includegraphics[scale=0.6]{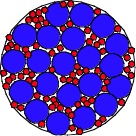}
\caption{(Color online) Schematic representation of a mixture in a volume $V$. Red and blue represent two different components in the mixture.}
\label{mix}
\end{figure}

The following describes the basic concept of our mix model. Let us first consider a volume $V$ 
containing the mixture of elements as depicted in the Fig.~\ref{mix} 
and suppose $N_k$ is the total number of atoms of an element $k$ in that volume. 
Assuming the volume $V$ can be divided into a collection of ion spheres, $v_k$,  
such that the spheres are the same size for all ions of the same element. 
Now $v_k$ is the quantity we wish to determine in the mixture. 

To begin, one must define the total number of atoms, total mass and total number densities according to
\begin{eqnarray}
N_{tot} = \sum_k N_k,
\end{eqnarray}
\begin{eqnarray}
M _{tot} = \sum_k M_k =  \sum_k N_k A_k,
\end{eqnarray}
\begin{eqnarray}
n_{tot} = \frac{N_{tot}}{V} = \sum_k \frac{N_k}{V} =  \sum_k n_k,
\end{eqnarray}
respectively, where $A_k$ is the atomic weight of element $k$, one can express the total mass density as
\begin{eqnarray}
\rho_{tot} = \sum_k \frac{N_kA_k }{V} = \sum_k \rho_k = \sum_k n_k A_k  = n_{tot} \sum_k x_k A_k = n_{tot} \bar A,
\label{rhotot}
\end{eqnarray}
where $x_k$ can be considered as the concentration of element $k$.

The quantity $\rho_k$ in Eq.(\ref{rhotot}) is the mass density for element $k$ in the mixture. Clearly, the $\rho_k$ of an element $k$ 
is not the same as if the whole volume $V$ contained only that element (it must be less). The effective density for element $k$ 
is defined by $\rho^*_k \equiv A_k/v_k$. Using the additive volume rule (i.e., the volume of a gas mixture 
is expressed as the sum of volumes occupied by the individual components with consideration the respective components 
to be at the pressure and temperature of the mixture) \cite{CRC}, one can obtain the relation between $\rho^*_k$ and $\rho_{tot}$ as 
\begin{eqnarray}
\frac{1}{\rho_{tot}} = \sum_k \left( \frac{x_kA_k}{\bar A}\right) \frac{1}{\rho^*_k}.
\label{mix1}  
\end{eqnarray}

Here, the additive volume rule is a constraint. It applies to all components in the mixture using an ion-sphere 
radius for each element. The set $\{\rho^*_k\}$ must satisfy this relation, but this constraint alone is insufficient to specify the individual values of $\rho^*_k$.  To proceed further, 
one must impose a physical condition connecting the atomic properties among the elements of the mixture. For example, we assume that the electronic pressure  
for each element in the mixture must be in equilibrium \cite{qeos}, that is to say,
\begin{eqnarray}
P_e = P_{e,k}(\rho^*_k,T)      
\end{eqnarray}
for each element $k$. This is the same assumption made in Thomas Fermi theory 
and requires that the set $\{\rho^*_k\}$ be determined iteratively based
on the condition of pressure equilibrium (e.g., $P_C = P_H$) and the constraint of additive volume rule. This procedure
has also been demonstrated for non-ideal gas equation of state \cite{PB2003}. 

The MIX algorithm uses the chemical potential equilibrium constraint instead of the pressure equilibrium, 
and Eq.(\ref{mix1}) is solved iteratively using the Brent algorithm \cite{NR}. In addition to the chemical potential 
equilibrium condition, there are other physical conditions like the electrical equilibrium obtained by 
ttreating the plasma as consisting of ions of different elements 
embedded in a sea of uniform free electrons within each ion-sphere. Similarly, one requires 
the electron density $n_e$ to be the equal for each element. Hence
the average charge state of each element $k$ is 
\begin{eqnarray}
n_e = \frac{\rm \bar Z_k(\rho^*_k,T)}{v_s} = \frac{ \rm \bar Z_k(\rho^*_k,T) \rho^*_k}{A_k}
\end{eqnarray}
This simple modification to Eq.(\ref{mix1}) is useful as it allows for the estimate of the 
non-LTE effects from the LTE calculation through the use of the RADIOM model 
\cite{BM2009}. After the partial mass densities for each element of the mixture are found,
they can be used to get the spectral opacity. It can be shown 
that the resulting opacity $\kappa_{mix}$ (in cm$^2$/g) of a mixture 
can be expressed as \cite{mix}
\begin{eqnarray}
\kappa_{mix}(\rho_{tot},T) = \frac{1}{\rho_{tot}}\sum_k \rho^*_k \kappa_k(\rho^*_k,T). 
\end{eqnarray}

\section{Results and Discussion}

\subsection{Comparison of theories and experiments for carbon plasma}

\begin{figure}[!htp]
\centering
\includegraphics[scale=0.1]{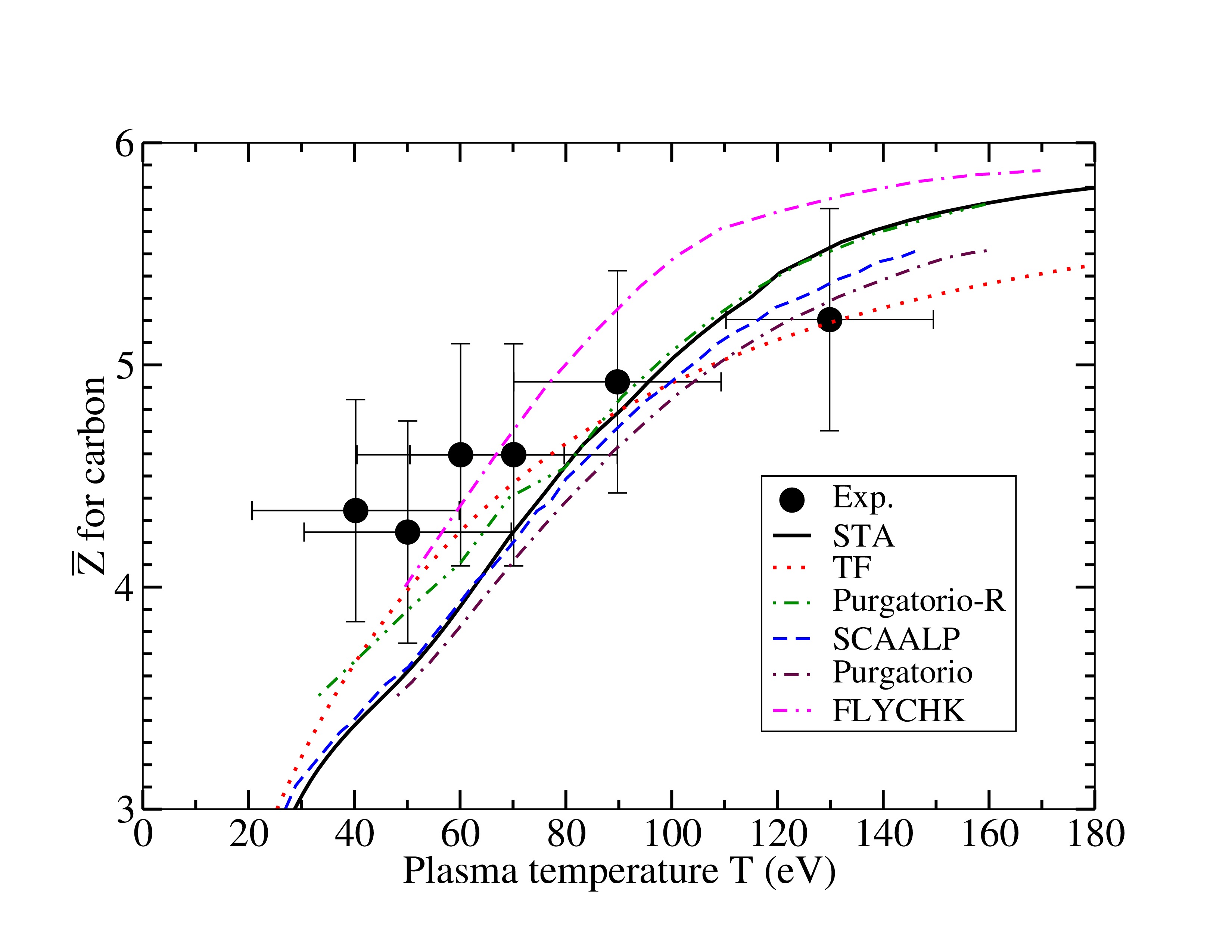}
\caption{(Color online) The carbon average ionization $\rm\bar Z$ as a function of temperature
at 0.2 g/cm$^3$ are compared to SCAALP \cite{SCAALP}, PURGATORIO \cite{Wilson}, 
FLYCHK \cite{flychk}, Thomas-Fermi (TF) model, and experimental data \cite{Gregori08}.}
\label{fig1}
\end{figure}

Spectrally-resolved X-ray scattering has been used to investigate the carbon average ionization 
in a multicomponent plasma in the high-temperature regime \cite{Gregori06, Sawada, Gregori08}. 
Let us first discuss the STA results for a carbon plasma. 
A good way to determine the quality of the STA calculations is by directly checking against experimental 
measurements rather than through the thermodynamic equation-of-state calculations or radiation-hydrodynamic simulations. 
This, however, is not always possible. Fig.~\ref{fig1} shows
the experimentally-determined average ionization of 
carbon at 0.2 g/cm$^3$ as a function of plasma temperature 
from 20 eV to 200 eV \cite{Gregori06, Sawada, Gregori08}. 
The STA results are compared to the benchmarked average ionization 
and temperature data obtained using spectrally and time-resolved X-ray scattering experiment at
the University of Rochester's OMEGA laser facility \cite{Gregori08}. 
It is shown that the STA calculated $\rm \bar Z$ values agree with the experimental 
data especially above plasma temperature of 70 eV, but underestimate them by about $30\%$ at 40 eV. 
In addition to the experimental data, we also consider theoretical results from SCAALP \cite{SCAALP}, PURGATORIO \cite{Wilson}, 
FLYCHK \cite{flychk} and Thomas-Fermi models. The SCAALP is based on neutral-pseudo-atom 
concept \cite{NPA1, NPA2} in which the model accounts for the effective ion-ion interaction
for any set of density and temperature parameters, starting from the self-consistent electron 
charge density previously calculated using density-functional theory for a single neutral pseudoatom. 
The general idea of the Purgatorio model, on the other hand, is based upon a spherically averaged 
ion embedded in jellium in a Dirac-Fock formulation \cite{Wilson}. Here, 
we display two sets of Purgatorio calculations, namely, the Purgatorio-R and 
Purgatorio. The Purgatorio-R is the results of the model with the continuum electrons in quasibound resonance states, 
whereas the Purgatorio is the results of the model without the continuum electrons in quasibound resonance states. 
Above the plasma temperature of 80 eV, we see that $\rm \bar Z$ coincides with the values predicted by Purgatorio-R,
but deviates from it below 80 eV to match up with the values predicted by SCAALP. 
Note that, like Purgatorio-R, the SCAALP model takes into account the continuum electrons in quasibound resonance states. 
This is unexpected since the STA model does not include the quasibound resonance effects. Another puzzling feature is that 
the the $\rm \bar Z$ from Purgatorio is in favorable agreement with results from the SCAALP model. 
We speculate that this kind of deviation could come from differences in the atomic orbital wave functions 
since they are rather sensitive to the particular central-field potential used to model the many-body interactions 
among electrons and ions in the atomic structure calculations. In any case, in agreement with SCAALP we find that the maximal 
principal quantum number is $n=3$ between around 40 and 160 eV. In addition, it is surprising the results from the simple 
TF model appeared to agree more closely with the experimental data. The present finding of the TF model also confirms 
the earlier TF result reported in Ref.\cite{SCAALP}. Finally, the results of the FLYCHK code \cite{flychk} 
closely agrees with the experimental data below a plasma temperature of 80 eV, but overestimate 
the experimental results at higher temperatures.  

\begin{figure}[!htp]
\centering
\includegraphics[scale=0.1]{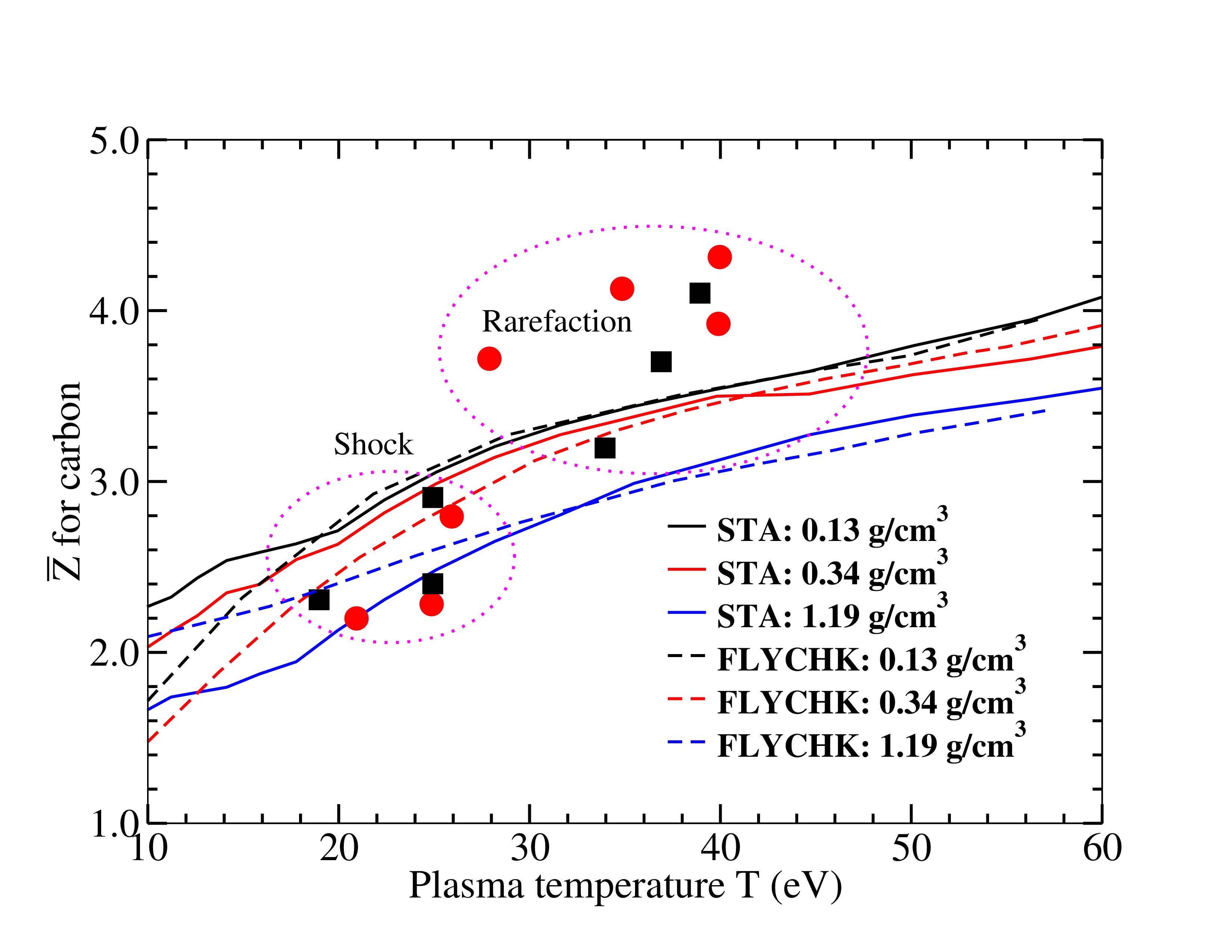}
\caption{(Color online) The carbon average ionization $\rm \bar Z$ as a function of plasma temperature.
The solid circles and squares are the measured $\rm\bar Z$ data from the
shocked foam \cite{MU}. The data are divided into the shocked layer and rarefaction groups.
The solid and dashed lines are the STA and FLYCHK results, respectively, evaluated at 2/5, 1, and 3.5 
times the initial or uncompressed carbon foam 
density of 0.34 g/cm$^3$. }
\label{fig2}
\end{figure}

\begin{figure}[!htp]
\centering
\includegraphics[scale=0.1]{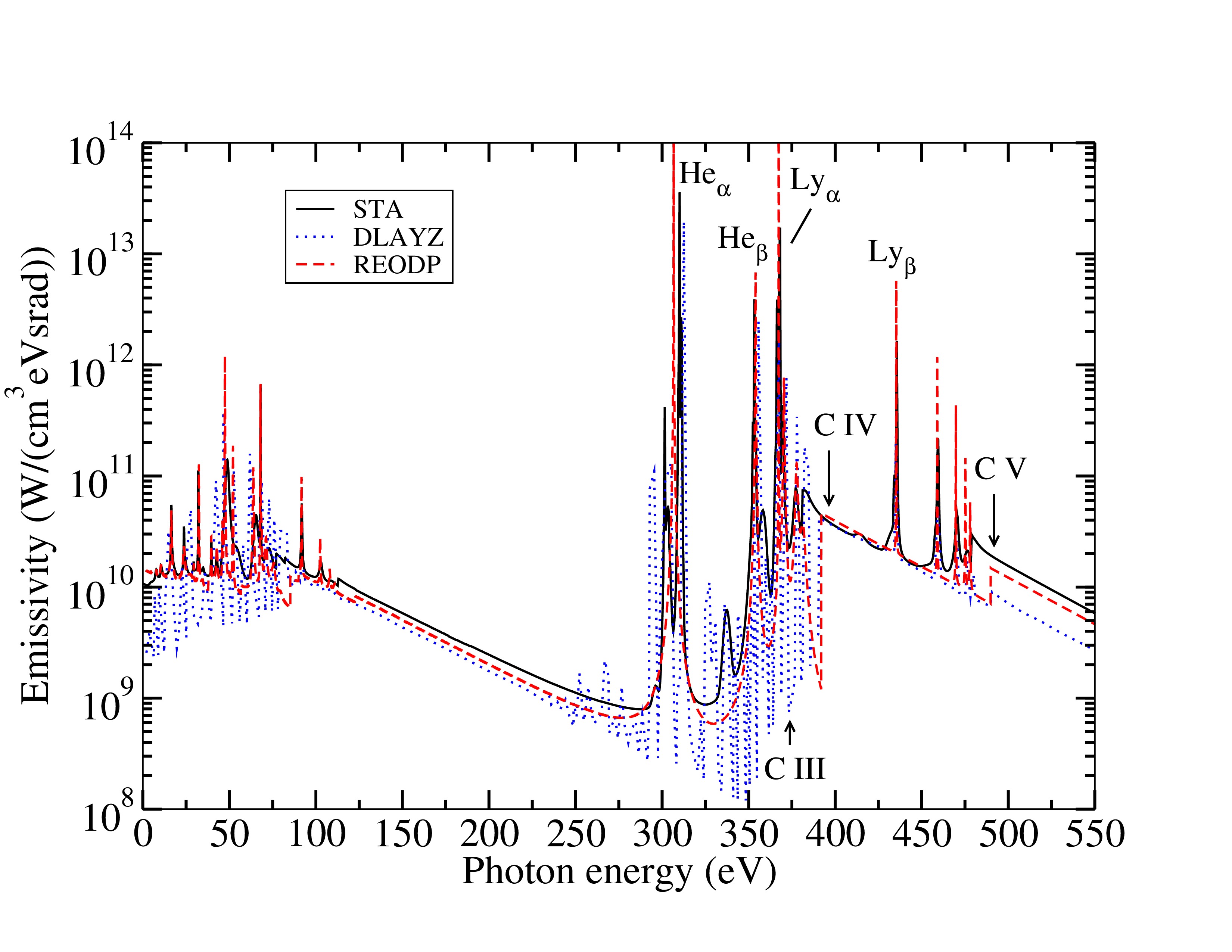}
\caption{(Color online) Emission spectrum of carbon plasma at a temperature
of 50 eV and density of 4.3 $\times ~10^{-3}$  g/cm$^3$. The locations of 1s 
ionization thresholds of C III, C IV and C V are indicated by arrows.}
\label{fig3}
\end{figure}

In Fig. \ref{fig2}, we consider the low temperature range of 20 - 60 eV and compute $\rm\bar Z$ for carbon. 
At such temperatures, the carbon ion is only moderately ionized. The STA result 
is compared with a more recent data from the University of Michigan and 
also with spatially-resolved X-ray scattering from the OMEGA facility \cite{MU}. It is obvious that, 
in the shocked region, $\rm \bar Z$ from STA at plasma density of 1.19 g/cm$^3$ 
agrees fairly well with the experimental data. However, in the rarefaction region, 
the STA predicts lower values of $\rm \bar Z$ than those suggested by the experiment. 
In comparison to the results from FLYCHK code \cite{flychk}, in general, the STA and FLYCHK results do show a similar
behavior of $\rm \bar Z$ as a function of temperature for all the densities except at temperature below 20 eV. 
Close examination in the region of shock layer, 
on average shows that, the discrepancy between the two calculations is about 10$\%$. 
In the region of rarefaction, on the other hand, both theoretical results agree more closely.

Next, we examine the emission spectrum of carbon plasma at a temperature
of 50 eV and the plasma density of 4.3 $\times ~10^{-3}$  g/cm$^3$. A comparison
of STA emission spectrum with those obtained using the non-LTE DLAYZ code \cite{Gao13}
and REODP code \cite{Gennady} is displayed in Fig. \ref{fig3}. 
The spectrum obtained using the DLAYZ code displays much finer details of the spectral lines. 
This is expected since the DLAYZ calculation is based on the 
Dirac-Fock-Slater DLA framework. The spectrum obtained using REODP code, on the other hand, 
is based on a less detailed Hartree-Fock-Slater approach. Nevertheless, the STA method reproduces 
many spectral features and their corresponding line locations displayed in the synthetic spectra obtained using 
the DLAYZ and REODP codes. From the STA calculations, we also find that, 
below the photon energy of $\sim$100 eV, the transition lines are attributed to the outer-shell 
electrons. For those localized in the energy range of 250 to 500 eV, they are from the inner-shell transitions. 
Following \cite{Gao13, Gennady}, we also have the locations of the 
1s ionization thresholds of C III, C IV, and C V ions, and the He-$\alpha$, He-$\beta$, 
Lyman-$\alpha$, and Lyman-$\beta$ emission lines marked in the figure. 
And from these markers, we notice that the STA calculated somewhat lower ionization edges for C IV and C V ions. 
These ionization edges appear to be shifted to lower thresholds by approximately 25 eV. 
A comparison of Fig.~4 with Fig.~5a and Fig.~6b suggests this shift may 
be caused by a {\em continuum lowering}, which is also known as an ionization potential depression 
(see page 322 of Ref. \cite{STA1}). Indeed, the ionization threshold given by the STA model, 
as shown in Fig.~6b, seems to converge with decreasing density 
to 480 eV given by the alternative DLAYZ and REODP models. Moreover, 
in the case of a higher density, displayed in Fig.~5a, the threshold shift predicted by 
the STA model has increased relative to Fig.~4, which is expected since
the position of the threshold in the alternative models suspiciously has not changed.
  
\begin{figure}[!htp]
\centering
\includegraphics[scale=0.1]{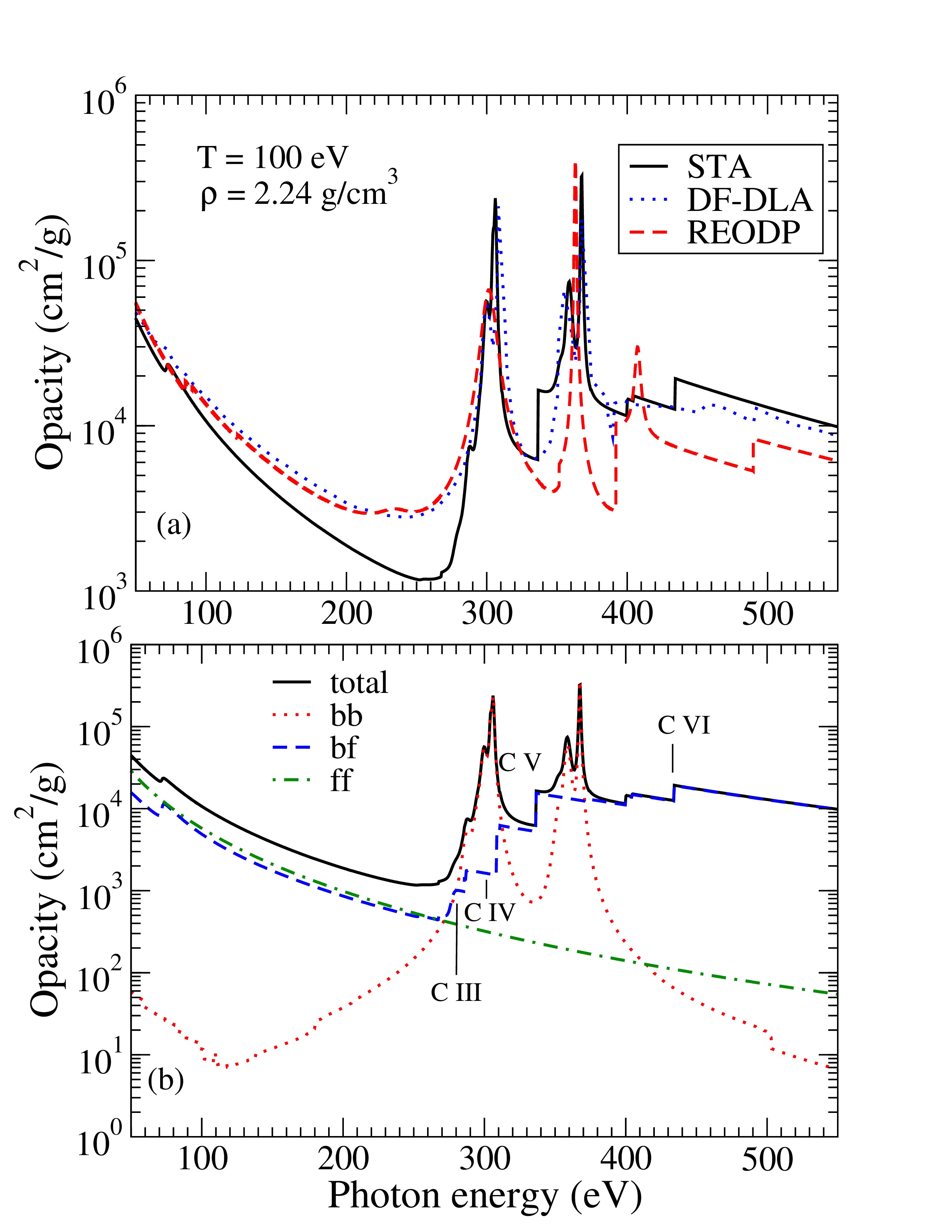}
\caption{(Color online)(a) A comparison of the radiative opacity (Planckian) of carbon at a density of 2.24 g/cm$^3$ 
and a temperature of 100 eV, (b) the breakdown of different contributions from the bound-bound, 
bound-free and free-free transitions of the total opacity.}
\label{fig4}
\end{figure}

Figure \ref{fig4}(a) shows the opacity of carbon plasmas 
at the density of 2.24 g/cm$^3$ and a temperature of 100 eV. The STA result is compared with 
the spectra calculated using another variant of Dirac-Fock-Slater DLA model (DF-DLA) 
by Gao {\it et al} \cite{Gao11} and the REODP code \cite{Gennady}. The STA-computed 
absorption spectra and line positions are reasonably consistent with respect to the results 
of the other two calculations, except below the photon energy of 280 eV where the STA-obtained 
opacities are much lower. The absorption structures found in the region of 0-120 eV result from 
transitions in the excited levels of carbon ions with the principal quantum number $n \ge$ 2. 
The absorption peaks near 300 and 380 eV, however, are the results of the 1s-2p transitions 
in C V and C VI, respectively, while other peaks at higher photon energy arise from the1s-$n$p ($n \ge$ 3) 
transitions of C VI. The breakdown of different contributions from the bound-bound, 
bound-free and free-free transitions which dictate the photon energy profile of the 
opacity is shown in Fig. \ref{fig4}(b). Below the photon energy of 280 eV, 
the bound-free and free-free transitions are dominant, and we see that STA 
predicts lower values of bound-free and free-free contributions in comparison to 
the DF-DLA and REODP calculations. Between 280 and 400 eV,  the bound-bound 
contribution becomes more important compared to the bound-free contribution. 
For the rest of the photon energy range, the opacity is dominated by the bound-free component. 
The opacity near the K-edge for different carbon ion stages is also noted on the bound-free curve. 

\begin{figure}[!htp]
\centering
\subfloat{\includegraphics[scale=0.1]{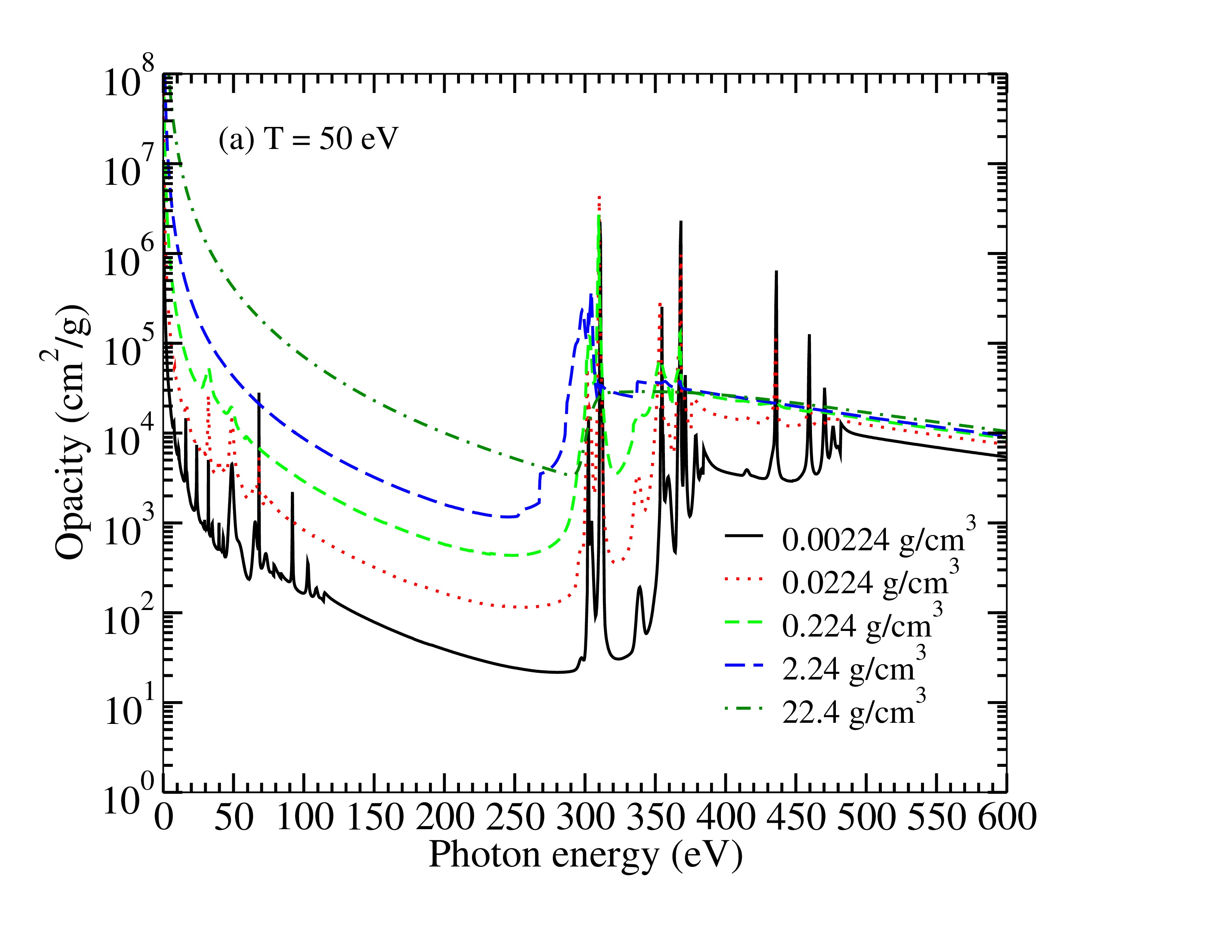}}
\vspace*{-0.8cm}
\subfloat{\includegraphics[scale=0.1]{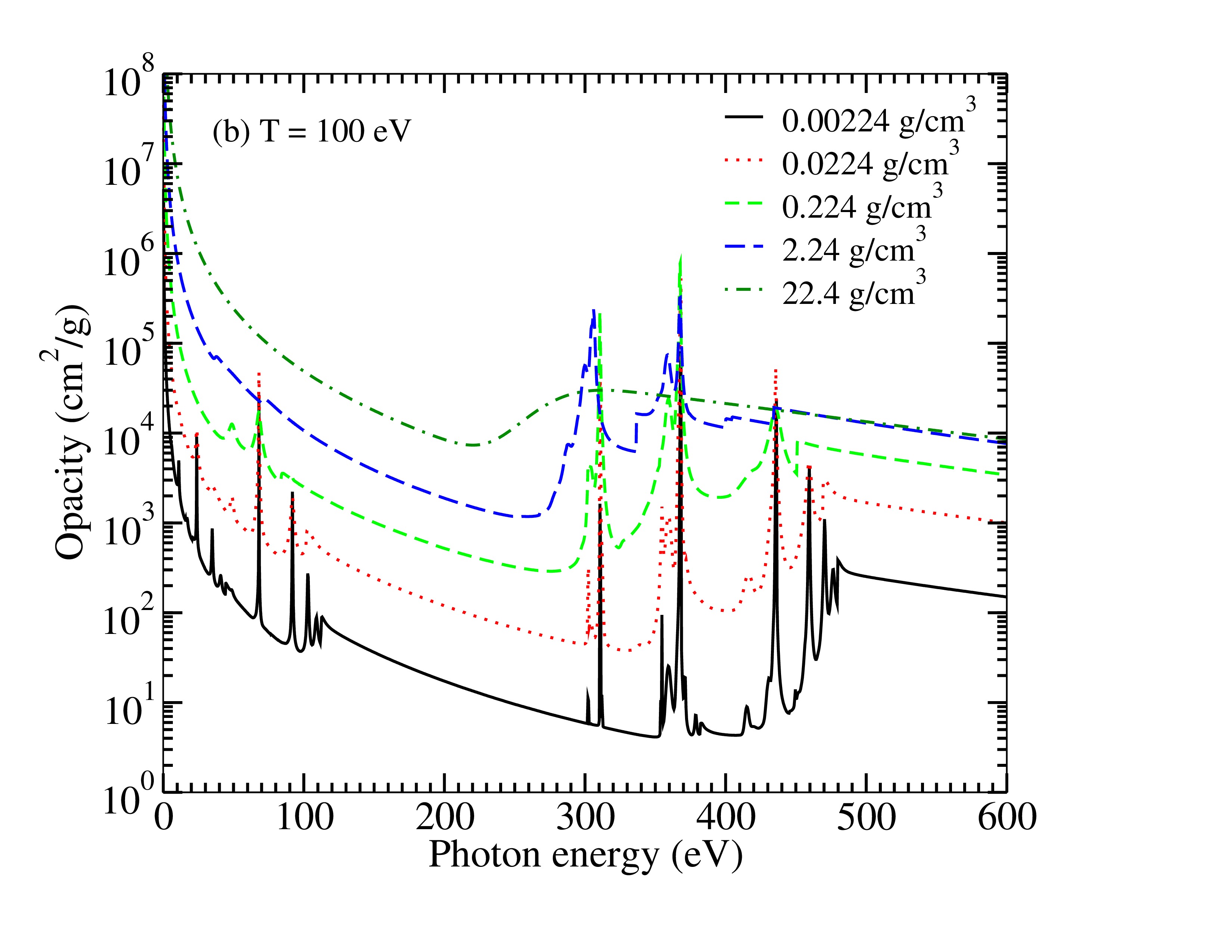}}
\caption{(Color online) The density dependence of the radiative opacity (Planckian) 
for carbon at a plasma temperature of (a) 50 eV and (b) 100 eV.}
\label{fig5}
\end{figure}

The dependence of the opacity on the plasma density of
carbon at 50 eV and 100 eV is shown in Fig. \ref{fig5}.  
Let us first discuss the plasma density effects. In this work, we restrict our attention 
to the ion-sphere model in which the STA method is based upon. Let us suppose that we have a plasma in an ion-sphere 
and the $N$-th ionization potential (IP) of an atomic system is the difference in binding energy ($\Delta E_B$) 
of the ground state of the atom ionized $N$ times and the ground state of that atom ionized ($N-1$) times. 
Classic atomic structure models, like Hartree-Fock methods, for example, treat
the atomic system as isolated such that no plasma effects are considered; consequently,  
a model such as the Stewart-Pyatt \cite{SP} or Ecker-Kr\"{o}ll \cite{EK} had to be introduced 
in order to account for the experimental fact that the $\Delta E_B$ depends strongly on the plasma density and temperature. 
Now, when considering an isolated atomic $N$-th ion, 
each bound electron is moving in the nuclear potential screened by the other electrons, 
the Coulombic potential as $r \to +\infty$ limit behaves like $V_C(r)\sim - (N-1)/r$.
In the STA model, and other methods (such as, INFERNO \cite{inferno}), 
the perturbation of the atom by the plasma (i.e., mostly by free electrons) is described by the Ion-Sphere (IS) model, 
the potential $V_{\rm IS}(r)$ vanishes at the boundary of the IS radius $R_{\rm IS} = (4\pi\rho_i/3)^{-1/3}$ 
with $\rho_i$ is the ionic density.  This is the ``charge neutrality condition'' used to obtain the balance 
between bound and free electrons within the ion sphere. As a consequence of this difference in potentials, 
the bound electrons in $V_{\rm IS}(r)$ are less bound than in $V_C(r)$. Therefore, the $\Delta E_B$ is 
smaller in a plasma than it is in an isolated atom.The phenomenon is termed the ``Ionization Potential Depression (IPD)". 
In the STA method, the IPD arises naturally for plasma conditions due to the combination of the IS model and the neutrality condition. 
Note that, in principle, the IS model should predict an IP lowering comparable to the high-density limit of the SP model, 
but often times the screening of the nucleus by bound electrons is model dependent. Figure~\ref{fig5} shows
the dependence of the opacity on the plasma density. Because of the depression of the IP as density rises,
the Inglis-Teller limit \cite{IT} where spectral lines merge together and show a quasi-edge appears to shift to 
lower and lower photon energies and gradually disappears into the continuum. 
Further increasing the density to hundred of times the solid density of carbon will gradually smear out
the Inglis-Teller limits, resulting in an exponentially attenuated function in photon energy 
for opacities at high-densities. A comparison of various IPD models, 
namely, the Stewart-Pyatt \cite{SP}, Ecker-Kr\"{o}ll \cite{EK}, modified Ion-Sphere \cite{MIS}, Crowley \cite{Crowley}, 
ATOMIC code \cite{atomic} with Hummer-Mihalas microfield ionization approach \cite{HM}, QMD 
and ``single mixture in a box" (SMIAB) models \cite{Hu2} has been discussed in the recent work of Hu \cite{Hu5} 
for the cases of temperatures near 10 eV. It is shown that the advanced ATOMIC, QMD and SMIAB 
codes predict an upward shift (in energy) of the carbon K-edge as the plasma density increases, while the others shows 
a downward shift. Note that, according to Hu \cite{Hu2}, the upward-shift of 
the carbon K-edge by high compressions can be attributed to the Fermi-surface energy rises 
faster in comparison to the continuum lowering process of the 1$s$-electron of carbon. This finding is somewhat surprising 
and warrants additional experimental confirmation. 

Figure.~\ref{fig5} also depicts the sensitivity of the opacities 
to the variation of the plasma temperature. At the lowest density of 0.00224 g/cm$^3$, 
comparing the opacities at temperatures of 50 eV and 100 eV, the opacity at 100 eV displays slightly 
fewer structures than those at 50 eV. This can be explained by observing that as plasma temperature increases, 
atoms become more ionized, and as a result, there are fewer bound electrons and thus simpler electronic configurations.

\subsection{Comparison of theories and experiments for CH plasma}

\begin{figure}[h]
\centering
\includegraphics[scale=0.1]{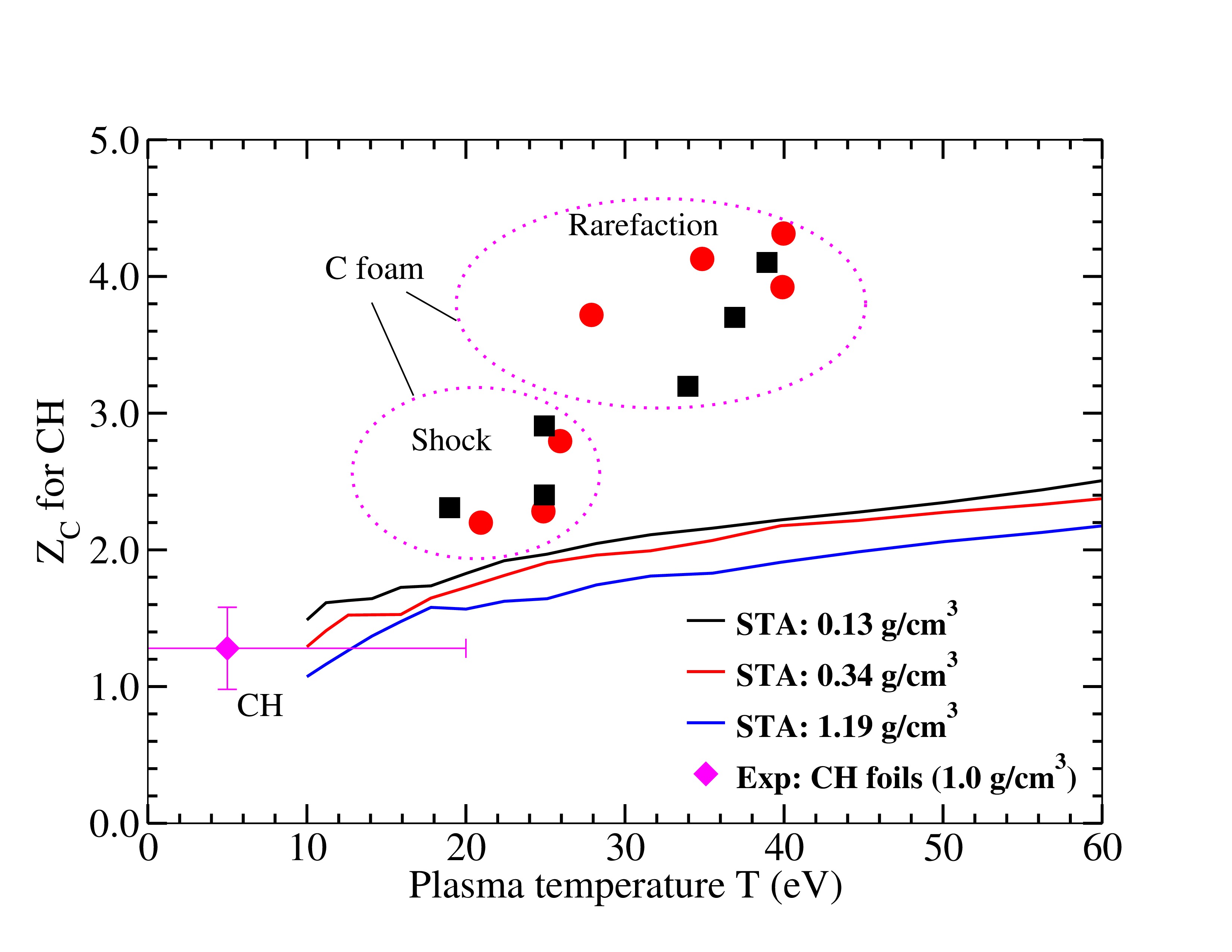}
\caption{(Color online) STA results for CH. 
The diamond is the experimental datum of CH foils from Fig.\ 8 
in Ref.\cite{Gregori06}.}
\label{fig6}
\end{figure}

It is informative to investigate how the average degree of ionization varies 
for a polystyrene (CH) target instead of a carbon foam. Figure.~\ref{fig6} shows 
the STA-predicted $\rm Z_C$ for CH. In the STA calculation, our polystyrene 
CH is a 1:1 ratio of C and H in mass density. It is shown 
that the calculated $\rm Z_C$ decreases to about 1.0 to 2.4 for CH, which is roughly 
50$\%$ smaller than the value for a pure carbon foam (see Fig.~\ref{fig2} for comparison) 
in the same temperature range. We also show the experimental 
datum \cite{Gregori06}, which is close to the STA prediction. 
We anticipate additional measurements of $\rm Z_C$ 
for a broader temperature range will be carried out in the near future as they are 
essential for validating theoretical calculations. 

\begin{figure}[h]
\centering
\includegraphics[width=9cm, height=12cm]{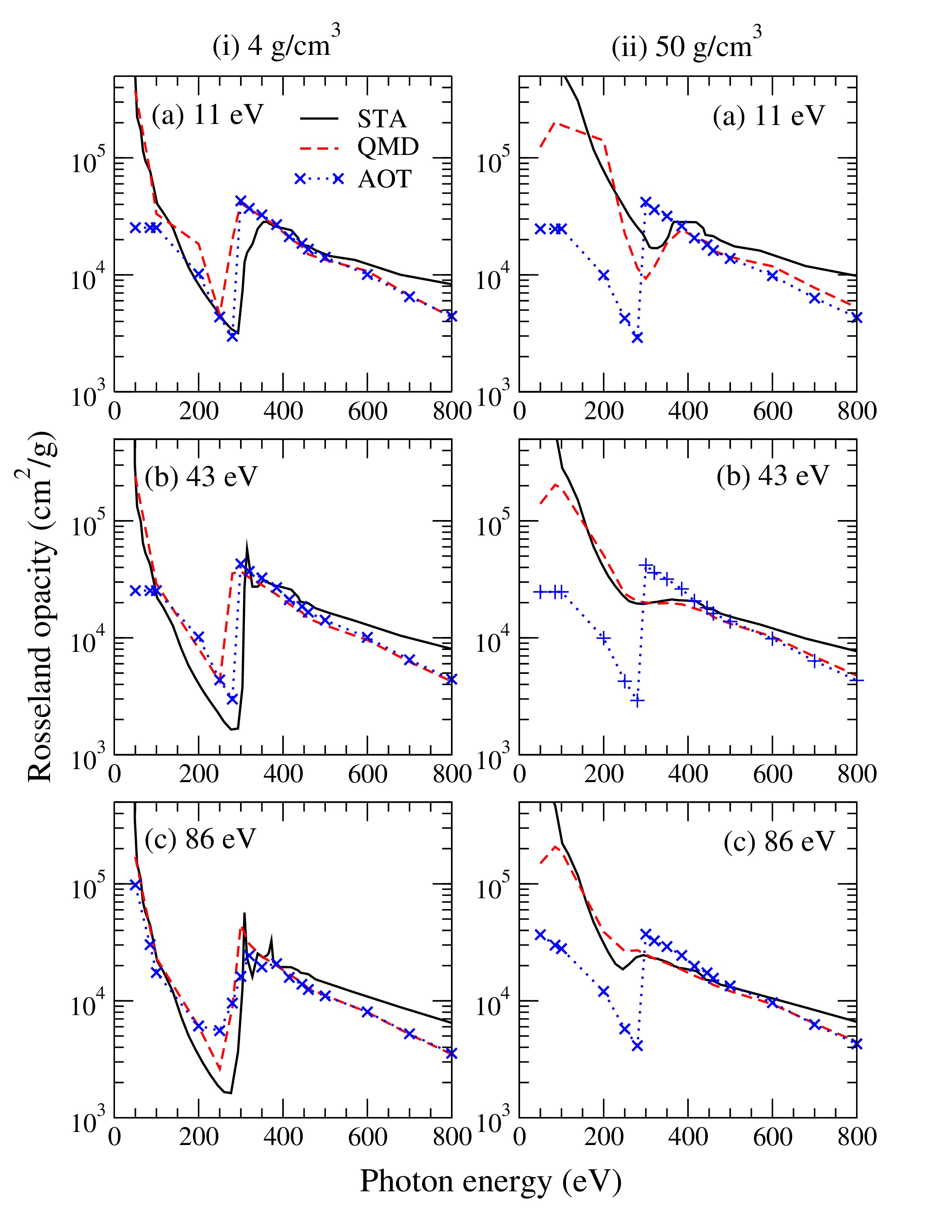}
\caption{(Color online)The Rosseland opacity of CH as a function of
photon energy for a plasma density of (i) 4.0 g/cm$^3$ and (ii) 50.0 g/cm$^3$, 
and temperature (a) 11.0 eV,  (b) 43.0 eV and (c) 86.0 eV.}
\label{fig7}
\end{figure}

Rosseland opacities have been calculated using the QMD method of Hu {\it et al} \cite{Hu2}. 
The QMD method is based on the finite-temperature density-functional theory (DFT). 
The many-body coupling and quantum degeneracy effects intrinsic to warm dense plasmas can be treated 
using the fundamental principles governing a quantum many-body system. For each QMD step, 
a set of electronic wave functions is self-consistently determined for a given ionic configuration. 
Then, the ions are moved classically with a velocity Verlet algorithm, 
according to the combined ionic and electronic forces. The ion temperature is kept constant through simple velocity scaling. 
Repeating these QMD steps results in a set of self-consistent ion trajectories and electronic wave functions. 
These trajectories provide a self-consistent set of static, dynamical, 
and optical properties in warm dense plasmas. In Fig. \ref{fig7}, we present the 
STA computed opacities and compare them to the results obtained 
using the QMD method at three plasma temperatures of 
11, 43 and 86 eV and densities of 4.0 g/cm$^3$ and 50.0 g/cm$^3$.  
For $\rho$ = 4.0 g/cm$^3$, there are notable differences
between the STA and the QMD calculations, particularly in the low-photon energy 
and high-photon energy regions. In the moderate photon energy range of 100-450 eV, 
the agreement is reasonable. Nevertheless, the STA results display a similar of 
photon energy dependence as the QMD method. For $\rho$ = 50.0 g/cm$^3$,
the STA computed opacities agree with QMD predictions between photon energies of 100 and 450 eV. 
Deviations appear for photon energies of below 100 eV and above 450 eV, and a difference 
of almost a factor of 2 is seen near 800 eV. Overall, the 
agreement between the STA and QMD results is encouraging for CH, since the STA method is 
known to be more accurate for high-Z material at high temperatures. In addition to the QMD results, 
we also consider the data from the AOT model \cite{AOT1}, since these data have been employed 
in the past to simulate ICF and HED experiments \cite{AOT2,AOT3,AOT4,AOT5}. 

\begin{figure}[h]
\begin{center}
\subfloat{\includegraphics[scale=0.1]{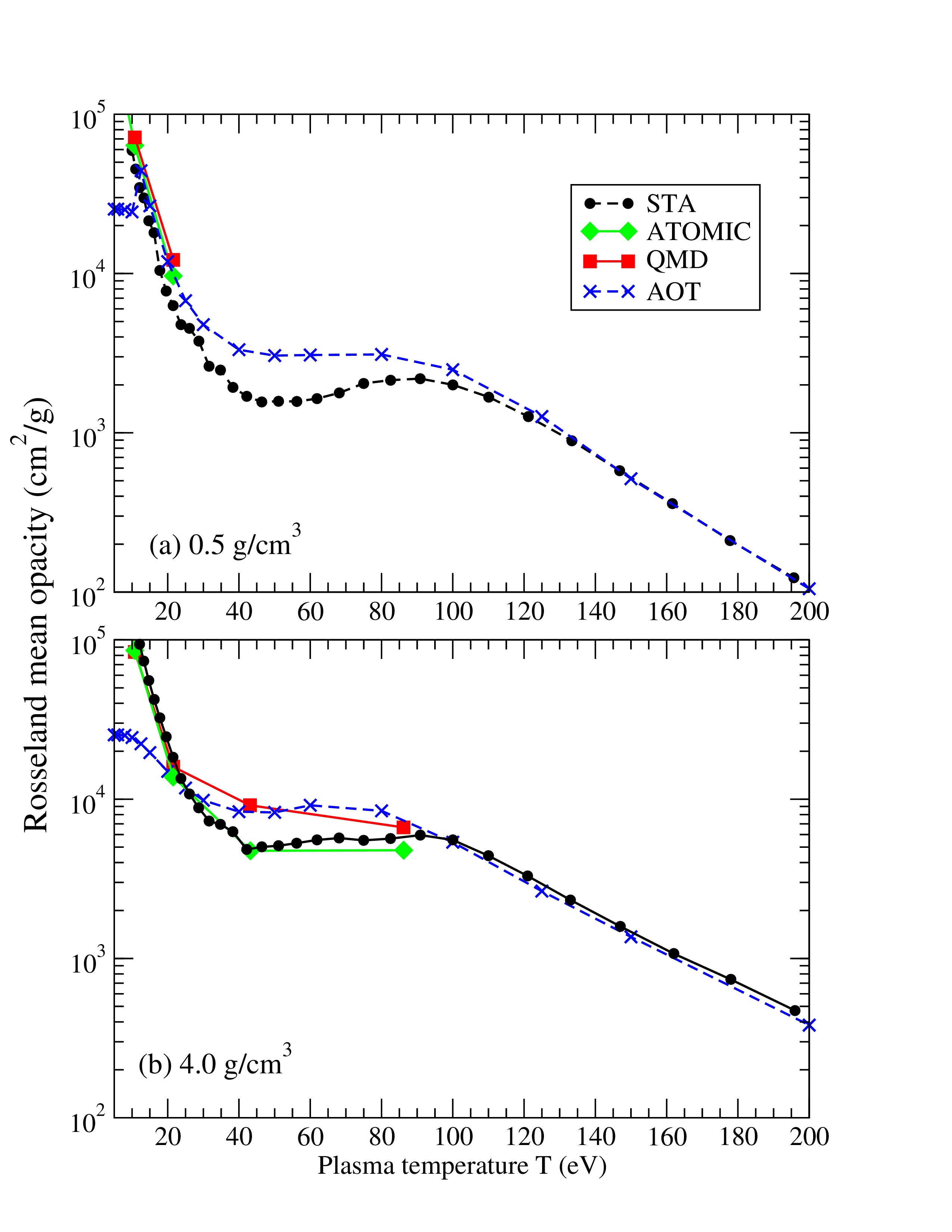}\label{fig:f8a}}
  \hspace*{-0.64cm}
\subfloat{\includegraphics[scale=0.1]{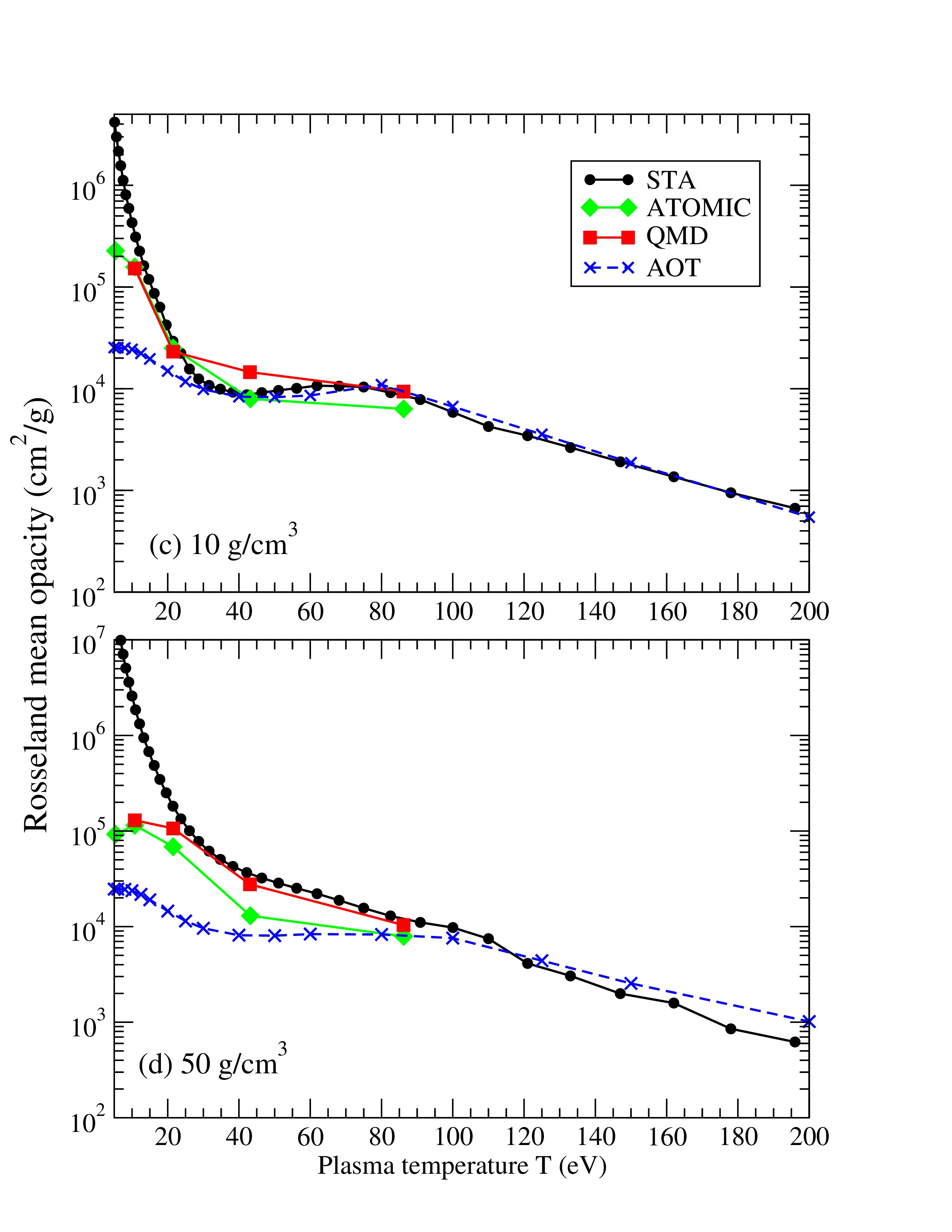}\label{fig:f8b}}
\caption{(Color online) The Rosseland mean opacity of CH as a function of
plasma temperature for four different densities: (a) 0.5 g/cm$^3$, (b) 4.0 g/cm$^3$, (c) 10.0 g/cm$^3$ 
and (d) 50.0 g/cm$^3$.}
\label{fig8}
\end{center}   
\end{figure}  

Finally, we examine the variations of Rosseland mean opacity for CH as a function of
plasma temperature for four different densities at 0.5 g/cm$^3$, 4.0 g/cm$^3$, 10.0 g/cm$^3$ 
and 50.0 g/cm$^3$. The results are shown in Fig.~\ref{fig8}. Along with the STA and QMD predictions, 
we also show the results from the ATOMIC model developed at the Los Alamos National Laboratory. 
The ATOMIC code is a suite of opacity-generation and kinetics codes \cite{atomic}. 
It's devised to generate opacities tables with very broad ranges of
temperatures and densities, and is usually used for
astrophysical modeling at low and moderate densities as well
as in radiation-hydrodynamics codes for the modeling of ICF
and of HED experiments. The newer version of ATOMIC code also takes
into account the IPD effects through plasma microfield ionization 
and the rising of the Fermi surface energy.

It is also of interest to compare the STA results with those from the ATOMIC code
and to the QMD method at high densities. As shown in Fig.~\ref{fig8}, we see the STA computed opacities 
are in reasonably good agreement with predictions from the ATOMIC (and QMD) calculations. At temperatures less than $\sim$20 eV,
the STA calculated opacities for 10.0 and 50.0 g/cm$^3$ deviate from the QMD and ATOMIC
results. In the low temperature region, both the QMD 
and ATOMIC calculations show a turning point near 10 eV, that approaches a finite value. 
The STA calculation, on the other hand, keeps rising and consequently overestimates the opacity in the sub-eV region. 
This limitation perhaps not surprising since the current STA model is not designed for low temperatures and high densities. 
It may be possible to circumvent this limitation by implementing the ``pressure-ionized-effective-statistical"
method of Busquet in the STA's PF algebra \cite{pies}. 

\section{Summary}
In summary,  we have used a super transition arrays (STA) 
method at local thermodynamics equilibrium to examine the emissivities, opacities and average ionization 
charge state for both carbon and CH plasmas with coupling constant $\Gamma$ varying 
from 0.02 to 2.0. Our objective is not only to validate the STA model, but also to assess 
the accuracy of the model by benchmarking the results of the  STA calculations 
against the available experimental data and results obtained using other theoretical methods. 
For carbon plasma, we find that STA reproduces emissivities 
and opacities with well-resolved spectral line features, including their corresponding locations,  
and are in good agreement with the results from Dirac-Fock and Hartree-Fock-Slater 
calculations. For CH plasmas, above temperatures of 20 eV, 
we find that STA-derived opacities agree reasonably well with the quantum molecular 
dynamics and Hartree-Fock calculations. 

Recently, x-ray scattering diagnostics of the blast wave in a planar carbon foam driven by 
the OMEGA laser have inferred the temperature of the carbon plasma is between 20 and 40 eV 
with $\rm\bar Z$ about 2.0 to 4.0 from the shock to rarefaction regions, respectively. 
Comparing with these experimental data, we find that the STA obtained $\rm Z_C$ 
is in agreement with the experiment in the shocked region, but is 
lower in the rarefaction region. The discrepancy
is consistent with the results reported using the FLYCHK code. Finally, 
we computed with STA the temperature dependence of  $\rm Z_C$ for CH 
in the same temperature range as for the carbon plasma. We find that 
these values vary from 1.0 to 2.0 from the shock to rarefaction regions, respectively. 
The present study shows the reliability of STA method for simulating 
low-Z multi-component WDM. Additional work based on the Busquet's ionization potential depression 
effective statistical weight approach is also in progress in order to improve the 
ionization potential depression model in the STA method.  
 
\vspace*{0.1cm}
\centerline {\bf Acknowledgment}
\vspace*{0.2cm} 
We thank Dr. Miloshevsky for sending us his theoretical results for comparison. 
The work is supported in part by the U.S. Department of Energy National Nuclear 
Security Administration.

\end{document}